\documentclass[fleqn,usenatbib]{mnras}

\usepackage{newtxtext,newtxmath}
\usepackage{float}
\usepackage{ulem}
\usepackage[T1]{fontenc}

\DeclareRobustCommand{\VAN}[3]{#2}
\let\VANthebibliography\thebibliography
\def\thebibliography{\DeclareRobustCommand{\VAN}[3]{##3}\VANthebibliography}

\usepackage{graphicx}	
\usepackage{amsmath}	

\title[VNC as a cosmological probe]{Void Number Counts as a Cosmological Probe for the Large-Scale Structure}

\author[Y. Song et al.]{
Yingxiao Song$^{1,2}$,
Qi Xiong$^{1,2}$,
Yan Gong$^{1,2,3}$\thanks{Email:gongyan@bao.ac.cn},
Furen Deng$^{1,2}$,
Kwan Chuen Chan$^{6,7}$,\newauthor
Xuelei Chen$^{1,2,4,5}$,
Qi Guo$^{1,2}$,
Yun Liu$^{1,2}$, 
and Wenxiang Pei$^{1,2}$
\\\\
$^{1}$National Astronomical Observatories, Chinese Academy of Sciences,20A Datun Road, Beijing 100012, China\\
$^{2}$School of Astronomy and Space Sciences, University of Chinese Academy of Sciences(UCAS),Yuquan Road NO.19A Beijing 100049, China\\
$^{3}$Science Center for China's Space Survey Telescope, National Astronomical Observatories, Chinese Academy of Sciences,\\20A Datun Road, Beijing 100101, China\\
$^{4}$Department of Physics, College of Sciences, Northeastern University, Shenyang 110819, China\\
$^{5}$Centre for High Energy Physics, Peking University, Beijing 100871, China\\
$^{6}$School of Physics and Astronomy, Sun Yat-sen University, 2 Daxue Road, Tangjia, Zhuhai, 519082, China\\
$^{7}$CSST Science Center for the Guangdong-Hongkong-Macau Greater Bay Area, SYSU, Zhuhai, 519082, China\\
}
\date{Accepted XXX. Received YYY; in original form ZZZ}

\pubyear{2015}

\begin{document}
\label{firstpage}
\pagerange{\pageref{firstpage}--\pageref{lastpage}}
\maketitle

\begin{abstract}

Void number counts (VNC) indicates the number of low-density regions in the large-scale structure (LSS) of the Universe, and we propose to use it as an effective cosmological probe. By generating the galaxy mock catalog based on Jiutian simulations and considering the spectroscopic survey strategy and instrumental design of the China Space Station Telescope (CSST), which can reach a magnitude limit $\sim$23 AB mag and  spectral resolution $R\gtrsim200$ with a sky coverage 17,500 deg$^2$, we identify voids using the watershed algorithm without any assumption of void shape, and obtain the mock void catalog and data of the VNC in six redshift bins from $z=0.3$ to1.3. We use the Markov Chain Monte Carlo (MCMC) method to constrain the cosmological and VNC parameters. The void linear underdensity threshold $\delta_{\rm v}$ in the theoretical model is set to be a free parameter at a given redshift to fit the VNC data and explore its redshift evolution. We find that, the VNC can correctly derive the cosmological information, and the constraint strength on the cosmological parameters is comparable to that from the void size function (VSF) method, which can reach a few percentage level in the CSST full spectroscopic survey. This is because that, since the VNC is not sensitive to void shape, the modified theoretical model can match the data better by integrating over void features, and more voids could be included in the VNC analysis by applying simpler selection criteria, which will improve the statistical significance. It indicates that the VNC can be an effective cosmological probe for exploring the LSS.
\end{abstract}


\section {introduction} \label{sec:intro}

Cosmic void, which is characterized by low density, large volume and linear evolution, is an important component of the cosmic large-scale structure (LSS). A large number of void samples can be obtained by galaxy surveys for studying the formation and evolution of the LSS and properties of dark energy and dark matter. 
A variety of cosmological probes associated with voids have been proven to be very effective, such as the galaxy-void cross-correlation in redshift space \citep[e.g.][]{2017JCAP...07..014H,2022A&A...658A..20H,correa2022redshift} and the void size function \citep[VSF, e.g.][]{pisani2015counting,verza2019void,2022A&A...667A.162C,contarini2023cosmological,2023MNRAS.522..152P,2023JCAP...12..044V,2024arXiv240114451V}. Cosmic voids are suitable for studying modified gravity and massive neutrinos due to their large volume and low density \citep[e.g.][]{cai2015testing,pisani2015counting,zivick2015using,pollina2016cosmic,achitouv2016testing,sahlen2016cluster,falck2018using,sahlen2018cluster,paillas2019santiago,perico2019cosmic,verza2019void,2019MNRAS.488.4413K,2019JCAP...12..055S,2021MNRAS.504.5021C,2022ApJ...935..100K,2023A&A...674A.185M,2023JCAP...12..044V,2023JCAP...08..010V}, and can also be used to measure baryonic acoustic oscillations (BAO) \citep[e.g.][]{chan2021volume,forero2022cosmic,khoraminezhad2022cosmic}. The great potential of voids has been shown in these existing studies, which can promote our research on the LSS of the Universe.

Here we propose to use void number counts (VNC) to explore the LSS and constraining the cosmological parameters. The VNC is the integral of the VSF over the void size at a given redshift. The VSF has been proven to be an effective method for constraining cosmological models in spectroscopic galaxy surveys, e.g. BOSS DR12 \citep{contarini2023cosmological}. As a function representing the number density of voids at different scales at a given redshift, the VSF can illustrate the features and evolution of voids. The theoretical models of the VSF are also developed in current relevant researches, that usually assume the voids have spherical shape. One of the popular theoretical VSF models is represented by the Sheth and van de Weygaert model \citep[SvdW,][]{sheth2004hierarchy}, and it assumes that large voids grow from isolated small voids. And the SvdW model was later extended to the volume conserving model \citep[$V{\rm d}n$,][]{jennings2013abundance}, which assumes that large voids merge from small voids rather than evolve independently, and it is now one of the most widely used VSF model.

However, the shape of voids can be arbitrary and irregular, and the current void finders are usually based on the watershed algorithm \citep{watershed}, without any assumption on the void shape. So it is necessary to trim the void catalogs for selecting the voids with spherical shape to match the theoretical model in current VSF studies, and recent works have trimmed void catalog in various degrees or improve existing theoretical models \citep[e.g.][]{contarini2023cosmological,2023MNRAS.522..152P,2023JCAP...12..044V,2024MNRAS.532.1049S,2024arXiv240114451V}. This obviously will exclude large numbers of voids, and dramatically lose statistical significance. On the other hand, if we consider the VNC, in principle, we do not need to trim the void catalog significantly, since the VNC is not as sensitive as the VSF to the void shape by integrating over the void size, and the theoretical model can be modified easily and may explain the data better. This will retain more voids in the analysis, and could improve the accuracy of cosmological constraint.

In order to investigate the feasibility of this method, we create mock galaxy and void catalogs based on simulations, and assume the China Space Station Telescope \citep[CSST,][]{zhan11,zhan2021csst,gong,2023MNRAS.519.1132M} as the instrument to perform the spectroscopic survey. We use the widely used watershed algorithm of the void finder and obtain the void effective radius, ellipticity, and volume weighted center needed in our analysis. Then we generate the mock data of the VNC in six redshift bins from $z=0.3$ to $1.3$, and  discuss two void samples with employing two selection criteria, i.e. applying empirical void size cut-off and the void ellipticity cut-off, respectively. We constrain the cosmological and void parameters by using the Markov Chain Monte Carlo (MCMC) method. We also compare the result to that from the VSF method given by \cite{2024MNRAS.532.1049S} based on the same galaxy catalog, and demonstrate the feasibility of the VNC method.

The paper is organized as follows: In Section \ref{sec:data}, we introduce the simulations for generating the mock galaxy and void catalogs; In Section \ref{sec:nv}, we discuss the computation of the theoretical model and generation of the mock data for the VNC; In Section \ref{sec:mcmc}, we show the constraint results of the cosmological and void parameters; The summary and conclusion is given in Section \ref{sec:discussion}.

\section{Mock Catalogs} \label{sec:data}

\subsection{Simulation} \label{sec:sim}

We use dark matter only N-body simulations, i.e. Jiutian simulations, to derive the mock galaxy and void catalogs. The Jiutian simulation we adopt covers a volume of $1~ (h^{-1}$Gpc)$^3$, and contains $6144^3$ particles with a mass resolution of $m_{\rm p}$ = $3.72 \times 10^8$ $h^{-1}M_\odot$. The simulation is run with the L-Gadget3 code, and uses friend-of-friend and subfind algorithm to identify dark matter halo and substructure \citep{2001NewA....6...79S,2005MNRAS.364.1105S}. It employs the fiducial cosmological model with $\Omega_{\text{m}} = 0.3111$, $\Omega_{\text{b}} = 0.0490$, $\Omega_\Lambda = 0.6899$, $n_{\text{s}} = 0.9665$, $\sigma_8 = 0.8102$ and $h = 0.6766$ \citep{2020A&A...641A...6P}. 
The redshift space distortion (RSD) and structure evolution effects are also considered, by constructing simulation cubes with slices from the outputting snapshots at different redshifts in the redshift range of a simulation box. In our mock catalog, we trace the merger tree of each galaxy and locate the snapshot with the closest redshift to the distance of the galaxy. Rather than directly slicing and stitching snapshots by redshift, our method avoids repetition and omission of galaxies at slice boundaries. We do not use interpolation to accurately calculate the RSD effect, as interpolating between snapshots does not capture galaxy position and velocity information accurately. This can only affect small scales in non-linear regime, and will not change our results at the scales we are interested in, since we exclude voids smaller than $5\ h^{-1}\text{Mpc}$ to avoid any impact on void identification and non-linear effect.

\subsection{Galaxy mack catalog} \label{sec:gcat}

We make use of the CSST spectroscopic galaxy survey as an example to construct the mock galaxy catalog. The CSST can simultaneously perform the photometric imaging and slitless spectroscopic surveys, covering 17500 deg$^2$ survey area and wavelength range 250-1000 nm. It has three spectroscopic bands, i.e. $\it GU$, $\it GV$ and $\it GI$, with the spectral resolution $R\gtrsim200$. The magnitude limit for a band can reach $\sim23$ AB mag for 5$\sigma$ point source detection. 

We construct the mock galaxy catalog based on an improved Semi-Analytic Model \citep{henriques2015galaxy}. The database contains luminosities of galaxy emission lines produced by post-processing as described in \cite{2024MNRAS.529.4958P}, which can be used to select galaxies detected by the CSST spectroscopic survey according to the signal-to-noise ratio (SNR) and measure the galaxy redshift. Here four emission lines, i.e. H$\alpha$, H$\beta$, [OIII] and [OII], are considered. Compared to hydrodynamical simulations, this semi-analytic model is good enough for our analysis, which can correctly produce the luminosity functions of the emission lines we are interested in. The SNR per spectral resolution unit can be estimated by \citep{cao2018testing,2022MNRAS.515.5894D}
\begin{equation}
    \text{SNR}=\frac{C_{\text{s}} t_{\text{exp}}\sqrt{N_{\text{exp}}}}{\sqrt{C_{\text{s}} t_{\text{exp}}+N_{\text{pix}}[(B_{\text{sky}}+B_{\text{det}})t_{\text{exp}}+R_{\text{n}}^2]}}\label{eq1},
\end{equation}
where $N_{\text{pix}}=\Delta A/l_{\text{p}}^2$ is the number of detector pixels covered by an object. Here $\Delta A$ is the pixel area on the detector, assumed to be the same for all galaxies for simplicity, and $l_{\text{p}} = 0.074''$ is the pixel size. The point-spread function (PSF) is assumed to be a 2D Gaussian distribution with the radius of 80\% energy concentration $\sim0.3''$ in the CSST spectroscopic survey. $N_{\text{exp}}$ is the number of exposures and $t_{\text{exp}}$ is the exposure time, and we set $t_{\text{exp}}=150\,\rm s$ and $N_{\text{exp}} = 4$. $R_{\text{n}}$ = 5 $e^-{\text{s}}^{-1}{\text{pixel}}^{-1}$ is the read noise, and $B_{\text{det}}$ = 0.02 $e^-{\text{s}}^{-1}{\text{pixel}}^{-1}$ is the dark current of the detector. $B_{\text{sky}}$ is the sky background in $e^-\text{s}^{-1}\text{pixel}^{-1}$, which is given by
\begin{equation}
    B_{\text{sky}} = A_{\text{eff}}\int I_{\text{sky}}(\nu)R_{\text{X}}(\nu)l_{\text{p}}^2\frac{d\nu}{h\nu},\label{eq2}
\end{equation}
where $A_{\text{eff}} = 3.14$ m$^2$ is the CSST effective aperture area, $I_{\text{sky}}$ is the surface brightness of the sky background, $R_X$ is the total throughput for band $X$ including filter intrinsic transmission, mirror efficiency, and detector quantum efficiency. Here we estimate $I_{\text{sky}}$ based on the measurements of earthshine and zodiacal light given in \cite{2012acs..rept....3U}. We find that $B_{\text{sky}}=0.016$, 0.196, and 0.266 $e^-\text{s}^{-1}\text{pixel}^{-1}$ for $GU$, $GV$ and $GI$ bands, respectively.  $C_{\text{s}}$ in Equation~(\ref{eq1}) is the counting rate from galaxy, and for emission line $i$ at frequency $\nu_i$, we have
\begin{equation}
    C_{\text{s}}^i = A_{\text{eff}}R_X\left(\frac{\nu_i}{z+1}\right)\frac{F_{\text{line}}^{i}}{h\nu_i/(z+1)},\label{eq3}
\end{equation}
where $F_{\text{line}}^{i}$ is the flux of the emission line $i$ which can be obtained from the simulation. 

We select galaxies if $\rm SNR\ge10$ for any emission line of the four lines H$\alpha$, H$\beta$, [OIII] and [OII] in any spectroscopic band to get the mock galaxy catalog. We find that the number density of galaxies are $\bar n =  1.5\times 10^{-2}, 4.3\times 10^{-3}, 1.2\times 10^{-3}, 4.6\times 10^{-4}, 2.0\times 10^{-4}$, and $9.0\times 10^{-5}$  $h^3{\rm Mpc}^{-3}$ for the six redshift bins at the central redshift $z_{\rm c} = [0.3,0.5,0.7,0.9,1.1,1.3]$, respectively. 
We also calculate the mean galaxy separation (MGS), which indicates the typical distance between two galaxies. It can be calculated by the number of galaxies $N_{\rm g}$ and the survey volume $V_{\rm s}$ at different redshift bins, i.e. MGS $=(V_{\rm s}/N_{\rm g})^{1/3}$, and the MGS value in each redshift bin is shown in Table~\ref{tab:1}. 
To include the RSD effect, the galaxy redshift is estimated by considering both galaxy peculiar motion and the accuracy of CSST slitless spectral calibration. The redshift $z$ is involving the peculiar motions of the source $z_{\text{pec}}$ and cosmological redshift $z_{\text{cos}}$ with relation $1+z = (1+z_{\text{cos}})(1+z_{\text{pec}}) = (1+z_{\text{cos}})(1+v_{\rm loc}/c)$, where $v_{\rm loc}$ is the LOS component of peculiar velocity. We also add a 0.2\% error to each redshift as the uncertainty of CSST slitless spectral calibration \citep{gong}.

\begin{table*}
    \center
    \caption{The number of voids and MGS with $R_{\rm v}>5 h^{-1}\text{Mpc}$ in the six redshift bins. The mean, minimum and maximum void radius with $R_{\rm v}>5 h^{-1}\text{Mpc}$, and the ranges of void radius and void numbers used in the analysis at different redshifts in Case~1 and Case~2 are also shown.\label{tab:1}}
    \begin{tabular}{cccccccc}
        \hline \hline
        Redshift &MGS& Void & R$_{\text{v}}^{\text{mean}}$ & R$_{\text{v}}^{\text{min}}$ & R$_{\text{v}}^{\text{max}}$& Radius range & Void number
        \\
        &&number&(Case 1/Case 2)&(Case 1/Case 2)&(Case 1/Case 2)&(Case 1/Case 2)&(Case 1/Case 2)
        \\
        &($h^{-1}$Mpc)&($>$5$h^{-1}$Mpc)&($h^{-1}$Mpc)&($h^{-1}$Mpc)&($h^{-1}$Mpc)&($h^{-1}$Mpc)\\
        \hline
        0.3&4.1&48016&11.0/11.8&5/5&45/45&(20,45)/(10,45)&2878/12234\\
        0.5&6.2&17966&16.1/17.5&5/5&62/62&(25,62)/(15,62)&2090/5222\\
        0.7&9.4&6500&23.4/25.2&5/7&77/77&(30,77)/(25,77)&1356/1617\\
        0.9&13.0&2882&31.4/33.7&7/11&90/90&(45,90)/(30,90)&393/868\\
        1.1&17.1&1322&39.5/42.1&11/17&148/148&(50,148)/(35,148)&285/475\\
        1.3&22.3&598&45.0/48.6&15/21&278/278&(70,278)/(45,278)&96/180\\		
        \hline
	\end{tabular}
\end{table*}

\subsection{Void mock catalog}

We identify voids in our mock galaxy catalog based on Voronoi tessellation and the watershed algorithm \citep{watershed}, without any assumption of void shape. The Void IDentification and Examination toolkit\footnote{\url{https://bitbucket.org/cosmicvoids/vide\_public/src/master/}} \citep[\texttt{VIDE},][]{vide} is chosen to find voids, which is based on ZOnes Bordering On Voidness \citep[\texttt{ZOBOV},][]{zobov}. The low-density zones found by the watershed algorithm are further merged in \texttt{VIDE}, and the merging condition is set to be the boundary between two adjacent low-density zones  $<$ 0.2 $\bar n$, where $\bar n$ is the average tracer density. To avoid void-in-void cases, we use the low-density zones that are not merged before as the void mock catalog.

\begin{figure}
	\includegraphics[width=\columnwidth]{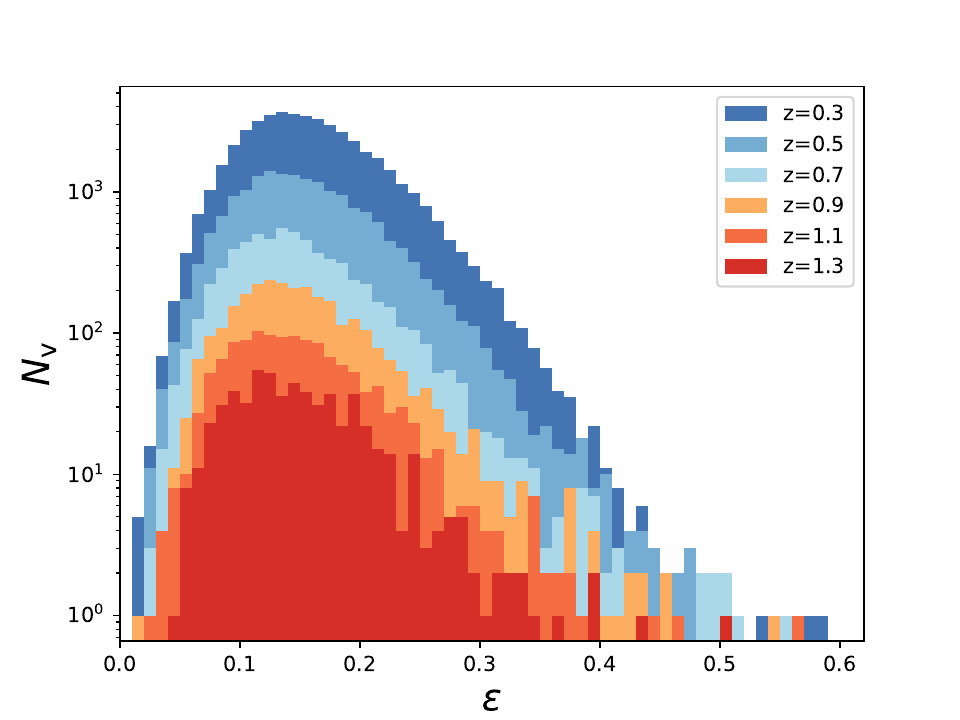}
    \caption{The void ellipticity distribution for voids with $R_{\rm v}>5 h^{-1}\text{Mpc}$ at different redshifts. The color from blue to red shows the void ellipticity distribution from $z$ = 0.3 to 1.3.}
    \label{fig:ved}
\end{figure}

The voids we obtain are composed of cells, and each cell contains a galaxy as the particle tracer. Based on the cell volume $V^i_{\text{cell}}$, we can compute the void volume $V$ and define the void effective radius $R_{\text{v}}$ in Eulerian space as
\begin{equation}
V=\sum V^i_{\rm cell}=\frac{4}{3}\pi R_{\rm v}^3.\label{eq4}
\end{equation}
The void volume-weighted center $\mathbf{X}_{\text{v}}$ also can be estimated by the cell positions as
\begin{equation}
\mathbf{X}_{\text{v}} = \frac{1}{V}\sum^N_i\mathbf{x}_i V^i_{\rm cell},\label{eq5}
\end{equation}
where $\mathbf{x}_i$ is the coordinate of the galaxy within a cell. 
We evaluate the void shape by estimating the ellipticity $\epsilon$, which can be derived by the smallest and largest eigenvalues of the inertia tensor, i.e. $J_1$ and $J_3$:
\begin{equation}
\epsilon = 1 - \left(\frac{J_1}{J_3}\right)^{1/4}.\label{eq6}
\end{equation}
The inertia tensor is given by
\begin{equation}
   I=\left[
 \begin{array}{ccc}
   I_{\rm xx} & I_{\rm xy} & I_{\rm xz}\\ 
   I_{\rm yx} & I_{\rm yy} & I_{\rm yz}\\ 
   I_{\rm zx} & I_{\rm zy} & I_{\rm zz}\\
 \end{array}
\right].    \label{eq7}
\end{equation}
The diagonal and off-diagonal components of the inertia tensor can be calculated as $I_{\text{xx}}= {\sum_{i=1}^{N}}(y^2_i+z^2_i)$ and $I_{\text{xy}} = -{\sum_{i=1}^{N}}x_iy_i$, where $x_i$, $y_i$ and $z_i$ denote the position of the galaxy in cell $i$ relative to the center of the void. To avoid the effect of nonlinear evolution, we only keep the voids with $R_{\rm v}>5\ h^{-1}\text{Mpc}$ in the analysis, which can retain as many as "usable" voids in our analysis \citep{2021MNRAS.500.4173S}. We also notice that this cut-off is only effective at low redshifts, since the minimum void radius $R_{\rm v}^{\rm min}$ is greater than $5\ h^{-1}\text{Mpc}$ at $z>0.7$ as shown in Table~\ref{tab:1}. We show the void ellipticity distribution from different redshift bins in Figure~\ref{fig:ved}. We can find that most voids in the mock catalog are spherical-like, and the peak of the void ellipticity distribution is about 0.15 at all redshift bins. Besides, the voids in higher redshift bins with larger $R_{\rm v}^{\rm mean}$ also have relatively lower ellipticity, i.e. more spherical-like voids reside at high redshifts.

Since the voids identified by the watershed algorithm can have arbitrary shapes, which will conflict with the void spherical evolution assumed in the theoretical model described in Section~\ref{sec:nv}, we need to make further void selection for model fitting. To reduce the discrepancy between the void data and theoretical model, we further trim our void catalog in two cases, i.e. Case~1 and Case~2, to explore the performance of the VNC method. Since larger voids are usually more spherical, in Case 1, we further select the voids only based on the void size, and empirically choose the void radius range considering the average void radius $R_{\rm v}^{\rm mean}$ at each redshift to retain large-size voids. We find that the lower limit of this range is about 2 times larger than $R_{\rm v}^{\rm mean}$ at $z=0.3$, and $\sim1.5$ times larger at $z>0.3$. The void radius ranges for different redshift bins are shown in Table~\ref{tab:1}. In Case~2, we basically choose the lower limit of the void radius range to include $R_{\rm v}^{\rm mean}$ (about 2.5$\times$MGS) in a redshift bin, and select voids with $\epsilon<0.15$ according to the peaks of the ellipticity distributions at different redshifts at $\epsilon \simeq0.15$. We find that selecting higher ellipticity cut-off, e.g. $\epsilon \sim0.2$, may not obtain spherical enough voids and lead to worse model fitting result. For Case 2, it has larger radius ranges and contains smaller voids compared to Case~1.

\section{Void Number Counts} \label{sec:nv}

Theoretically, the VNC are obtained by integrating the VSF over the void radius in a survey volume $V_{\rm S}$ at a given redshift, and it can be calculated by
\begin{equation}\label{eq8}
N_{\text{v}}(z)=V_{\rm S}\int_{R_{\text{v}}^{\text{min}}}^{R_{\text{v}}^{\text{max}}}\frac{dn}{dR_{\text{v}}}dR_{\text{v}}.
\end{equation}
By assuming that the volume fraction $V{\rm d}n$ is constant from Lagrangian space to Eulerian space \citep{jennings2013abundance}, the VSF can be estimated as
\begin{equation}
\frac{{\rm d}n}{{\rm d}R_{\text{v}}}=\frac{\mathcal{F}(\nu)}{V(R_{\rm v})}\frac{R_{\text{L}}}{R_{\text{v}}}\frac{{\rm d}\nu}{{\rm d}R_{\text{L}}},\label{eq9}
\end{equation}
where $R_{\text{L}}$ is the Lagrangian void radius. The peak height $\nu$ is decided by the void linear underdensity threshold of void formation $\delta_{\text{v}}$ and the root-mean-squared density fluctuation at different redshifts, which is given by $\nu=|\delta_{\text{v}}|/\sigma_M(z)$,
where $\sigma_M(z)=\sigma_0(R_{\rm L})D(z)$ and $D(z)$ is the linear growth factor. We obtain $\sigma_M(z)$ using \texttt{CAMB} in our work \citep{camb}. For the void radius ranges we choose, $\mathcal{F}(\nu)$ can be approximated very well by \citep{sheth2004hierarchy} 
\begin{equation}
\mathcal{F}(\nu)=\sqrt{\frac{2}{\pi}}\,{\rm exp}(-\frac{\nu^2}{2})\,{\rm exp}(-\frac{|\delta_{\text{v}}|}{\delta_{\text{c}}}\frac{\mathcal{D}^2}{4\nu^2}-2\frac{\mathcal{D}^4}{\nu^4}).\label{eq10}
\end{equation}
Here $\mathcal{D}=|\delta_{\text{v}}|/(\delta_{\text{c}}+|\delta_{\text{v}}|)$, and $\delta_{\text{c}}=1.686$. $\delta_{\rm v}$ usually can be derived theoretically assuming spherical evolution, and it is found to be $-2.731$ in the $\Lambda$CDM model when the void matter desity $\rho_{\rm v}=0.2\bar{\rho}_{\rm m}$ \citep{jennings2013abundance}. However, since the voids identified in our catalog do not assume any shape, we set $\delta_{\rm v}$ as a free parameter at a given redshift in the model fitting process, and it can be seen as an effective void linear underdensity threshold in our model. Then the relation between $R_{\text{L}}$ and $R_{\text{v}}$ is given by
\begin{equation}\label{eq:RL}
R_{\text{L}}\simeq\frac{R_{\text{v}}}{(1-\delta_{\text{v}}/c_{\rm v})^{c_{\rm v}/3}},
\end{equation}
where $c_{\rm v} =1.594$ \citep{bernardeau1993nonlinear,jennings2013abundance}.

In addition, we also consider the RSD and Alcock-Paczy\'{n}ski \citep[AP,][]{1979Natur.281..358A} effects in the model by two factors, for relating the void radius in observational redshift and real spaces. Following \cite{correa2021redshift}, we have
\begin{equation}\label{eq12}
R_{\rm v}^{\rm obs} = f_{\rm RSD}f_{\rm AP}R_{\rm v}.
\end{equation}
Here $f_{\rm RSD} = 1 -(1/6)\beta\Delta(R_{\rm v})$, where $\Delta(R_{\rm v}) = (1-\delta_{\rm v}/c_{\rm v})^{-c_{\rm v}}-1$ \citep{bernardeau1993nonlinear}, $\beta = f/b$, $f$ is the growth rate, and $b$ is the tracer bias. We set $\beta$ as free parameters at different redshift bins in the fitting process. $f_{\rm AP}=\alpha_{\parallel}^{1/3}\alpha_{\perp}^{2/3}$, where $\alpha_{\parallel} = H^{\rm fid}(z)/H(z)$ and $ \alpha_{\perp} = D_{\rm A}(z)/ D_{\rm A}^{\rm fid}(z)$, which are the ratios for  Hubble parameters and angular diameter distances in the fiducial and real cosmologies. 

\begin{figure} 
   \centering
   \includegraphics[width=\columnwidth]{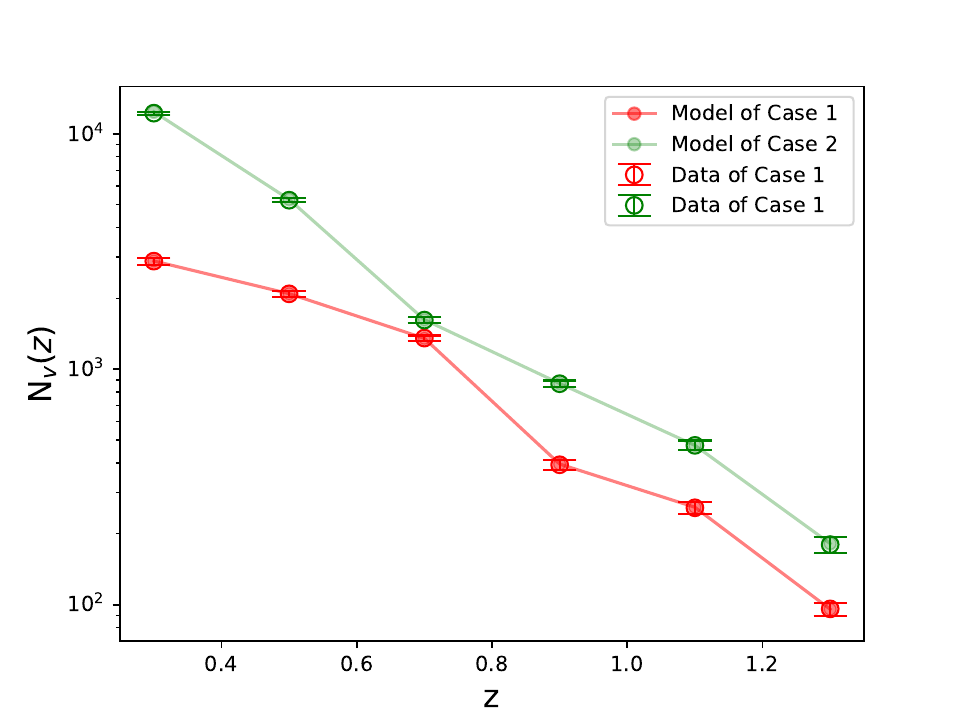}
   \caption{The void number counts of Case~1 (red) and Case~2 (green) for the six redshift bins from z = 0.3 to 1.3. The curves are the best-fits of the theoretical model, and the points with error bars denote the mock VNC data.}
   \label{fig:void count}
   \end{figure}

In Figure~\ref{fig:void count}, we show the VNC mock data and best-fit theoretical curves for Case~1 (red) and Case~2 (green). We use jackknife method to derive the error bar of each data point. We can find that, as expected, the number of voids decreases as the redshift increases, and the void number in Case~1 is lower than Case~2 since more voids are excluded in Case~1 by applying narrower void radius ranges.

\section{CONSTRAINT and result} \label{sec:mcmc}
We adopt $\chi^2$ for constraining the model parameters, which takes the form as
\begin{equation}\label{eq13}
\chi^2 = \sum_{z\, {\rm bins}}\left[\frac{N_{\text{v}}^{\text{data}}(z_i)-N_{\text{v}}^{\text{th}}(z_i)}{\sigma^i_{\text{v}}}\right]^2,
\end{equation}
where $N_{\text{v}}^{\text{data}}(z_i)$ and $N_{\text{v}}^{\text{th}}(z_i)$ are the data and theoretical model of the VNC, respectively, and $\sigma^i_{\text{v}}$ is the data error in the $i$th redshift bin. The likelihood function can be calculated by $\mathcal{L}$ $\propto$ exp($-\chi^2$/2).

\begin{table}
    \center
    \caption{The best-fit values, errors, and relative accuracies of the six cosmological parameters, the void linear underdensity threshold parameter $\delta_{\rm v}$ and the RSD parameter $\beta$ in the six redshift bins from $z$ = 0.3 to 1.3 for the two cases have been shown. We should note that the constraint results will be improved by about one order of magnitude, if considering the full CSST spectroscopic galaxy survey with 17500 deg$^2$ survey area.
    \label{tab:2}}
    \renewcommand{\arraystretch}{1.5}
    \begin{tabular}{ccccccc}
        \hline  \hline
        Parameter & Best-fit value of Case 1&
        Best-fit value of Case 2
        \\
        \hline
        $w$& $-1.386_{-0.300}^{+0.427}(26.2\%)$&$-1.307_{-0.326}^{+0.394}(27.5\%)$\\
        $h$&  $0.714_{-0.073}^{+0.070}(10.0\%)$&$0.768_{-0.091}^{+0.079}(11.1\%)$\\
        $\Omega_{\text{m}}$&  $0.282_{-0.082}^{+0.129}(37.5\%)$&$0.324_{-0.081}^{+0.111}(29.6\%)$\\
        $\Omega_{\text{b}}$&$0.036_{-0.012}^{+0.025}(51.2\%)$&$0.043_{-0.016}^{+0.023}(45.4\%)$\\
        $n_{\text{s}}$& $0.897_{-0.143}^{+0.199}(19.0\%)$&$0.951_{-0.174}^{+0.163}(17.7\%)$\\
        $A_{\text{s}}(\times 10^{-9})$& $2.518_{-0.557}^{+0.356}(18.1\%)$&$2.416_{-0.608}^{+0.422}(21.3\%)$\\
        \hline
        $\delta_{\text{v}}^1$& $-0.598_{-0.145}^{+0.140}(23.8\%)$&$-1.424_{-0.314}^{+0.354}(23.5\%)$\\
        $\delta_{\text{v}}^2$& $-0.288_{-0.091}^{+0.098}(32.8\%)$&$-0.703_{-0.154}^{+0.141}(21.0\%)$\\
        $\delta_{\text{v}}^3$&$-0.182_{-0.071}^{+0.050}(33.2\%)$&$-0.229_{-0.131}^{+0.073}(44.5\%)$\\
        $\delta_{\text{v}}^4$ &$-0.109_{-0.067}^{+0.038}(48.2\%)$&$-0.213_{-0.083}^{+0.062}(34.0\%)$\\
        $\delta_{\text{v}}^5$&$-0.131_{-0.048}^{+0.042}(34.1\%)$&$-0.189_{-0.068}^{+0.044}(29.9\%)$\\
        $\delta_{\text{v}}^6$&$-0.079_{-0.035}^{+0.030}(40.7\%)$&$-0.148_{-0.053}^{+0.036}(30.1\%)$\\
        \hline
        $\beta^1$& $0.264_{-0.176}^{+0.162}(63.8\%)$&$0.225_{-0.155}^{+0.178}(74.0\%)$\\
        $\beta^2$& $0.259_{-0.177}^{+0.171}(67.2\%)$&$0.285_{-0.188}^{+0.148}(59.0\%)$\\
        $\beta^3$& $0.293_{-0.189}^{+0.145}(57.0\%)$&$0.278_{-0.186}^{+0.157}(61.6\%)$\\
        $\beta^4$&$0.270_{-0.183}^{+0.159}(63.4\%)$&$0.257_{-0.168}^{+0.174}(66.5\%)$\\
        $\beta^5$&$0.259_{-0.175}^{+0.162}(65.1\%)$&$0.271_{-0.185}^{+0.162}(64.0\%)$\\
        $\beta^6$& $0.258_{-0.174}^{+0.167}(65.8\%)$&$0.250_{-0.165}^{+0.172}(67.3\%)$\\		
        \hline
	\end{tabular}
\end{table}

\begin{figure*}
	\includegraphics[scale=0.5]{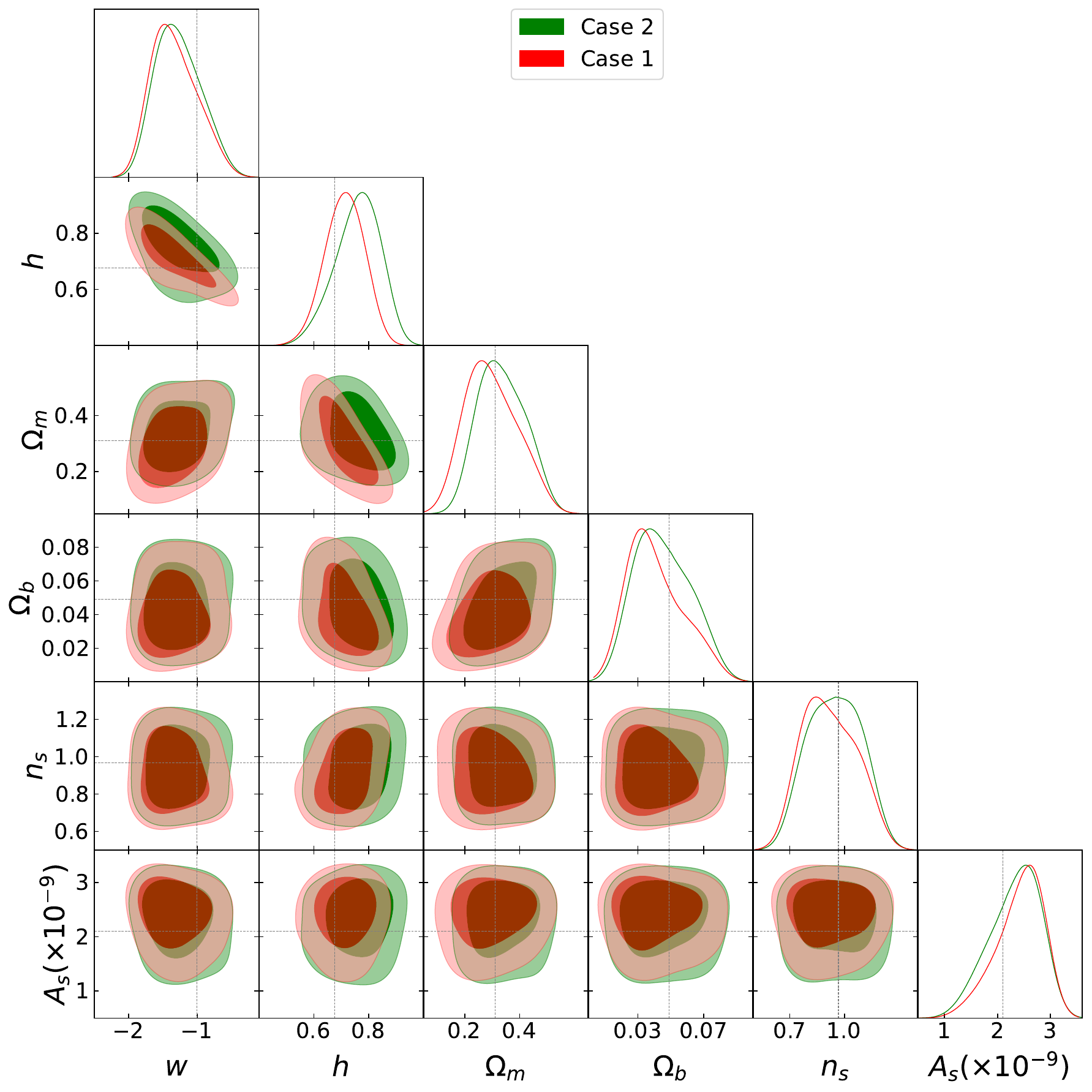}
    \caption{1D PDFs and contour maps of the cosmological parameters at 68\% and 95\% CL from the VNC for Case~1 (red) and Case~2 (green). The gray vertical and horizontal dotted lines denote the fiducial values of the parameters.}
    \label{fig:mcmccosmic}
\end{figure*}

We constrain the free parameters using the Markov Chain Monte Carlo (MCMC) method with \texttt{emcee} \citep{emcee}. We choose 112 walkers and obtain 20000 steps for each chain. The first 10 percent of steps are discarded as the burn-in process. In our model we totally have 18 free parameters. The cosmological free parameters are the dark energy equation of state  $w$, reduced Hubble constant h, the total matter density parameter $\Omega_{\text{m}}$, baryon density parameter $\Omega_{\rm b}$, spectral index $n_{\rm s}$, and amplitude of initial power spectrum $A_{\rm s}$. The free parameters about void $\delta_{\text{v}}^i$ is the threshold for void formation from the six redshift bins from $z=0.3$ to 1.3. And we also set the RSD parameter $\beta^i$ in the six redshift bins. The flat priors of the free parameters are $\Omega_{\rm m}\in(0.1,0.5)$, $\Omega_{\rm b}\in(0.02,0.08)$, $A_{\text{s}}/10^{-9}\in(1.0,3.0)$, $n_{\rm s}\in(0.7, 1.2)$, $w\in(-1.8,-0.2)$, $h\in(0.5,0.9)$, and $\delta_{\text{v}}^i\in(-2,0)$ and $\beta^i\in(0, 0.5)$ in the six redshift bins. And we set the fiducial value of cosmological parameter with $w=-1$, $h = 0.6766$, $\Omega_{\text{m}} = 0.3111$, $\Omega_{\text{b}} = 0.049$, $n_{\text{s}} = 0.9665$, and $A_{\rm s}/10^9=2.1$. Note that $\delta_{\text{v}}^i$ and $\beta^i$ are not the input parameters in the simulation, and they do not have the fiducial values.

In Figure \ref{fig:mcmccosmic}, we show the one-dimensional (1D) probability distribution functions (PDFs) and contour maps of the six cosmological parameters constrained by the VNC for Case~1 and Case~2 at 68\% and 95\% confidence levels (CL). The details of the constraint result of all the free parameters are shown in Table \ref{tab:2}, containing the best-fit values, 1$\sigma$ errors, and relative accuracies. We can find that the fitting results of the cosmological parameters are consistent with the corresponding fiducial values within 1$\sigma$ CL. This means that our method of the VNC can derive the cosmological information correctly.

Besides, the constraint powers in Case~1 and Case~2 are similar for the cosmological parameters, which give the constraint accuracies as $\Omega_{\rm m}\sim30\%$, $\Omega_{\rm b}\sim50\%$, $A_{\text{s}}/10^{-9}\sim20\%$, $n_{\rm s}\sim20\%$, $w\sim27\%$, and $h\sim10\%$. Note that the constraint accuracy can be improved by about one order of magnitude, if considering the full CSST spectroscopic survey covering 17500 deg$^2$. We can find that this result is comparable to that using the VSF method based on the same simulation and galaxy catalog as shown in \cite{2024MNRAS.532.1049S}. This is probably because that the method of the VNC is not as sensitive as the VSF method to the void shape. Hence, we can adopt simpler selection criteria, keep more voids in the cosmological analysis and include more redshift bins, i.e. six redshift bins from $z=0.3$ to 1.3 in this work while only four bins from $z=0.5$ to 1.1 in \cite{2024MNRAS.532.1049S}, which can effectively improve the statistical significance. Besides, our theoretical modeling without fixing $\delta_{\rm v}$ also could match the data better, since detailed void features are integrated over in the VNC. All these advantages can make the VNC obtain stringent and accurate constraints on the cosmological parameters. Here we do not show the constraint result of $\beta$, since we find that the VNC is not very sensitive to $\beta$ and there is no stringent constraint on it (with an accuracy $>60\%$) for the current mock data we use.

\begin{figure}
	\includegraphics[width=\columnwidth]{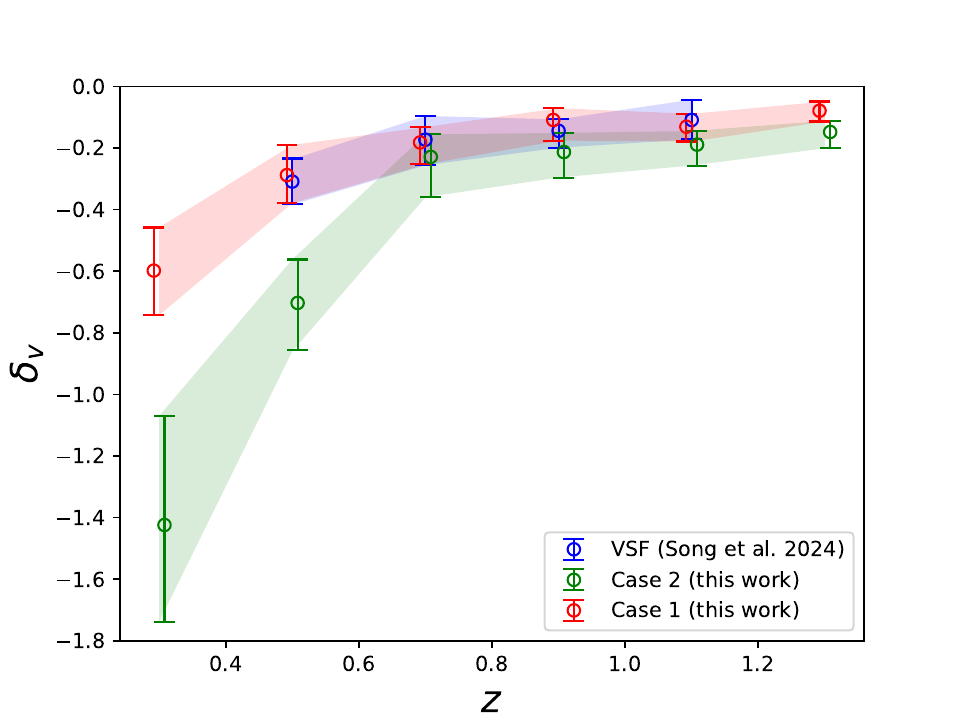}
    \caption{The best-fit values and 1$\sigma$ errors of $\delta_{\text{v}}$ in the six redshift bins from $z=0.3$ to 1.3 for Case~1 (red) and Case~2 (green). For comparison, the results from the VSF method in four redshift bins from $z=0.5$ to 1.1 are also shown in blue data points \citep{2024MNRAS.532.1049S}.}
    \label{fig:dvcompare}
\end{figure}

In Figure~\ref{fig:dvcompare}, we plot the best-fit values and 1$\sigma$ errors of $\delta_{\rm v}$ in the six redshift bins for Case~1 and Case~2, and the results from the VSF method given by \cite{2024MNRAS.532.1049S} are also shown for comparison. The contour maps and and 1D PDFs of $\delta_{\rm v}^i$ are shown in Appendix~\ref{sec:appendix}. We can find that the values of $\delta_{\rm v}$ in all the three cases have significant discrepancy compared to the theoretical values $\delta_{\rm v}\simeq-2.7$, assuming the spherical evolution and simulation particles as tracers \citep{jennings2013abundance}. Since our voids are identified by the watershed algorithm without any assumption of void shape and adopting galaxies as tracers, the current result is reasonable.

We also notice that the values of $\delta_{\rm v}$ at different redshifts in the three cases have a similar trend, which are approaching to 0 when redshift increases. This means that, as expected in the linear evolution, the Eulerian void size $R_{\rm v}$ will be close to the Lagrangian void size $R_{\rm L}$ at high redshifts (see Eq.~(\ref{eq:RL})). In addition, the result of Case~1 is more consistent with that from the VSF method in $z=0.5-1.1$, since the two analyses select similar ranges of void radius. On the other hand, the values of $\delta_{\rm v}$ in Case~2 are always lower than the other two cases, especially at $z<0.6$. This is because the value of $\delta_{\rm v}$ can reflect the merging of small voids into large voids in the $V{\rm d}n$ model. Typically, a smaller $\delta_{\rm v}$ means the average void radius in a sample is smaller \citep[see e.g.][]{2013MNRAS.436.3525R}. Since the lower limit of the void radius range in Case~2 is smaller than that in Case~1, Case~2 contains more small voids used in the analysis. This will suppress the average void radius in Case~2 and lead to a smaller $\delta_{\rm v}$ compared to Case 1, especially at low redshifts. If choosing similar radius ranges for Case~1 and Case~2, we find the values of $\delta_{\rm v}$ will be closer in these two cases.

\section{Summary and conclusion}
\label{sec:discussion}

In this work, we propose to use the VNC as a cosmological probe for studying the LSS. To check the feasibility, we generate the galaxy mock catalog based on Jiutian simulations and the CSST spectroscopic galaxy survey, and identify voids by $\tt VIDE$ without assuming void shape. The mock void catalog and data of the VNC are then derived at the six redshift bins from $z=0.3$ to $1.3$ in two cases, i.e. Case~1 and Case~2, by using empirical void radius ranges and considering the void ellipticity, respectively. We also set $\delta_{\rm v}$ as a free parameter at a given redshift in the theoretical model for better fitting the mock data and studying its redshift dependency and evolution. The RSD and AP effects are also considered in the analysis. Then we perform a joint fit of the cosmological, void and RSD parameters using the mock VNC data by the MCMC method.

We find that both Case~1 and Case~2 with different selection criteria lead to similar results. For the constraints on the cosmological parameters, the VNC can correctly derive the cosmological information, that the constraint power is comparable to the VSF, and can provide a few percentage level constraints on the cosmological parameters in the CSST spectroscopic survey. This is due to that the VNC is insensitive to void shape, and more voids and redshift bins can be kept by simpler selection criteria  in the analysis, which could effectively improve the statistical significance. The theoretical model of the VNC also can be effectively modified by assuming a free $\delta_{\rm v}$ at a given redshift, and can match the data better than the VSF by integrating over the void size. For the constraint on the void linear underdensity threshold $\delta_{\rm v}$, the results at different redshift bins from the VNC method have a similar trend as that from the VSF method, i.e. it becomes larger and larger and close to zero when redshift increases. The value of $\delta_{\rm v}$ is also dependent on the chosen void radius range at a given redshift. All of these indicate that the VNC can be a feasible and effective probe in cosmological studies.

\section*{Acknowledgements}

YS and YG acknowledge the support from National Key R\&D Program of China grant Nos. 2022YFF0503404, 2020SKA0110402, and the CAS Project for Young Scientists in Basic Research (No. YSBR-092). KCC acknowledges the support the National Science Foundation of China under the grant number 12273121. XLC acknowledges the support of the National Natural Science Foundation of China through Grant Nos. 11473044 and 11973047, and the Chinese Academy of Science grants ZDKYYQ20200008, QYZDJ-SSW-SLH017, XDB 23040100, and XDA15020200. QG acknowledges the support from the National Natural Science Foundation of China (NSFC No.12033008). This work is also supported by science research grants from the China Manned Space Project with Grant Nos. CMS- CSST-2021-B01 and CMS-CSST-2021-A01.

\section*{Data Availability}

 The data that support the findings of this study are available from the corresponding author upon reasonable request.



\bibliographystyle{mnras}
\bibliography{nvref} 




\appendix
\section{MCMC results for $\delta_{\rm v}$}\label{sec:appendix}

\begin{figure}
	\includegraphics[scale=0.28]{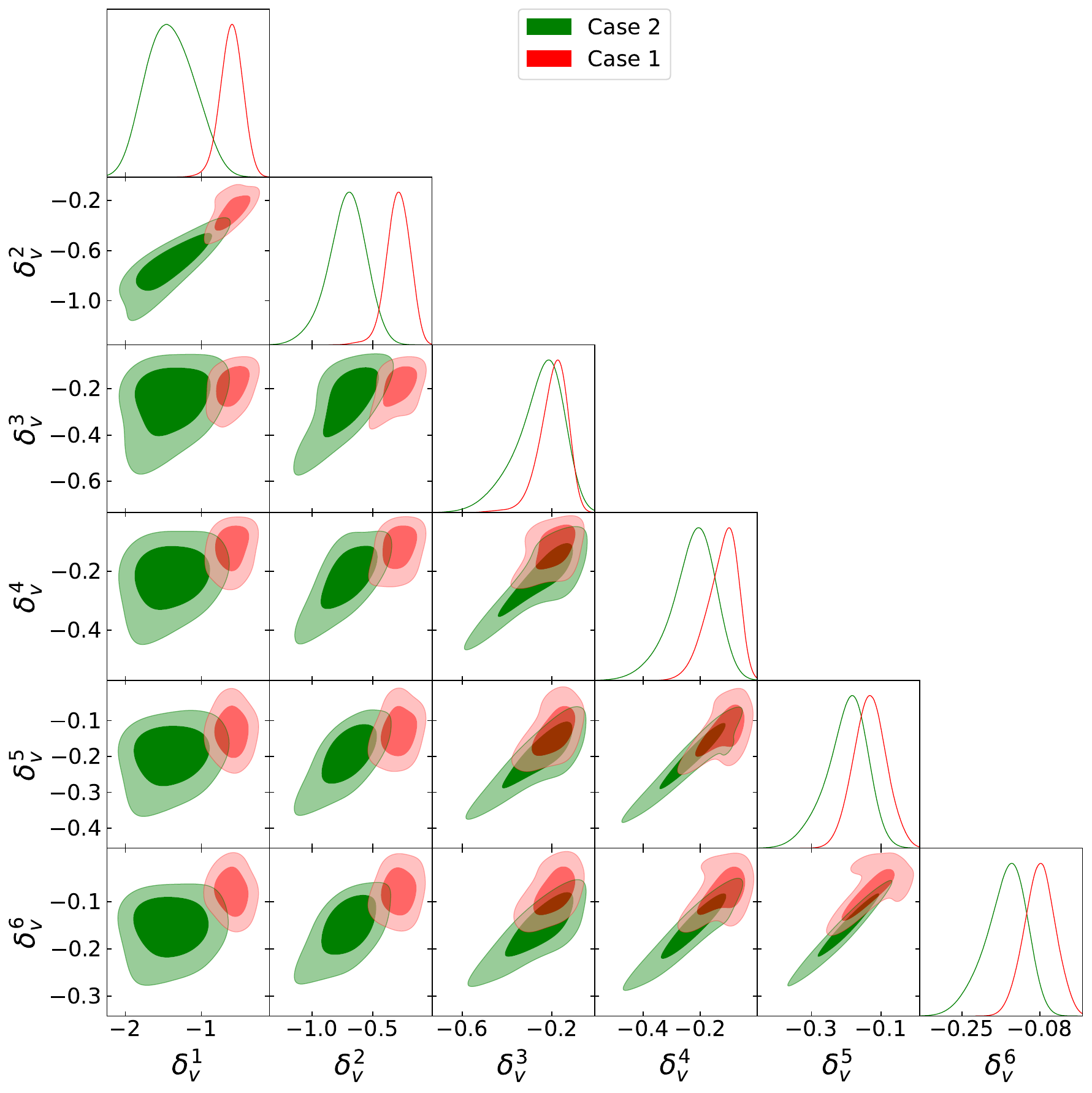}
    \caption{The 1D PDFs and contour maps at 68\% and 95\% CL of the linear underdensity thresholds for void formation $\delta_{\text{v}}^i$ derived from the VNC method in the six redshift bins from $z=0.3$ to 1.3. The red and green colors denote the results for Case~1 and Case~2, respectively.}
    \label{fig:mcmcdv}
\end{figure}

In Figure~\ref{fig:mcmcdv}, we show the contours at 68\% and 95\% CL and 1D PDFs of the linear underdensity thresholds for void formation $\delta_{\text{v}}^i$ in the six redshift bins for Case~1 (red) and Case~2 (green). Because $\delta_{\text{v}}^i$ is not the input parameter in the simulation, it does not have the fiducial value. The details of the best-fit values, 1$\sigma$ errors, and relative accuracies for the six $\delta_{\text{v}}^i$ are also shown in Table \ref{tab:2}.


\bsp	
\label{lastpage}
\end{document}